\documentclass[12pt,a4paper]{article}
\usepackage[utf8]{inputenc}
\usepackage{amsmath}
\usepackage{physics}
\usepackage{amsfonts}
\usepackage{amssymb}
\usepackage{makeidx}
\usepackage{xcolor}
\usepackage{graphicx}
\usepackage{lmodern}
\usepackage{float}
\usepackage[left=2cm,right=2cm,top=2cm,bottom=2cm]{geometry}

\title{Compactification, T-Duality and Quantum Erasers}
\date{}
\begin{document}
\maketitle
\vspace{-9ex}
\begin{center}

\author{\large{Salman Sajad Wani$^1$, Dylan Sutherland$^2$, Behnam Pourhassan$^{3,5}$,} \\ \large{Mir Faizal$^{2,4,5}$, Hrishikesh Patel$^6$}

\vspace{2ex}

  \textit{\small$^1$Department of Physics, University of Kashmir, Srinagar, Kashmir, 190006  India}\\
  \textit{\small$^2$ Department of Physics and Astronomy, University of Lethbridge,}\\
  \textit{\small Lethbridge, AB T1K 3M4, Canada}\\
  \textit{\small$^3$School of Physics, Damghan University, Damghan, 3671641167, Iran} \\ 
   \textit{\small$^4$Irving. K. Barber School of Arts and Sciences, University of British Columbia,} \\
  \textit{\small Okanagan Campus, Kelowna, V1V1V7, Canada}\\
  \textit{\small$^5$Canadian Quantum Research Center,
204-3002 32 Ave Vernon, BC V1T 2L7  Canada}\\
  \textit{\small$^6$Department of Physics and Astronomy, University of British Columbia,} \\
  \textit{\small6224 Agricultural Road, Vancouver, V6T 1Z1, Canada}

  }
\end{center}
  
\begin{abstract}
 \noindent Using T-duality, we will argue that a zero point length exists in the low energy effective field theory of string theory on compactified extra dimensions. Furthermore, if we neglect all the oscillator modes, this zero point length would  modify low quantum mechanical systems. As this zero length is fixed geometrically, it is important to analyze how it modifies purely quantum mechanical effects. Thus, we will analyze its effects on quantum erasers, because they are based on quantum effects like entanglement. It will be observed that the behavior of these quantum erasers gets modified by this zero point length. As the zero point length is fixed by the radius of compactification, we argue that this results demonstrate a deeper connection between geometry and quantum effects.  
 \end{abstract}

 \maketitle
 
\section{Introduction}
It is expected that due to quantum gravitational effects, the Planck length will act as a zero point length in spacetime     \cite{a, a1, a2, b1, b2}, and this will remove  ultraviolet divergences  in   quantum field theories
  \cite{b4, b5}. Thus, an ultraviolet completion of quantum field theories from zero point length would occur  due to quantum gravitational  corrections \cite{a, a1, a2, b1, b2}. In fact, it is known that string theory is one of the best candidates for quantum gravity, and string theory is an ultraviolet finite theory. This ultraviolet finiteness of string theory occurs due to a zero point length  in perturbative string theory \cite{c1, c2}. 
This is because the fundamental string is the smallest probe in perturbative string theory, and so the spacetime 
cannot be probed below the string length scale. In fact, it has been demonstrated that 
in perturbative string theory, this zero point  length $l_{0}$ is given by $l_{0} = g_s^{1/4} l_s$ (where $l_s = \alpha'$ is the string length, and $g_s$ is the string coupling constant). Even though the non-perturbative point like objects, such as D0-branes are present in non-perturbative string theory, it has been demonstrated that a minimal length of the order of 
$l_{0} = l_s g_s^{1/3}$ also exists in non-perturbative string theory \cite{a2, s18}. Such a minimal length occurs due to T-duality of string theory, as it can be argued  using  T-duality that the description of string theory 
above the string length scale $(l_s)$ is the same as its description below string length scale. Thus, the string length scale acts as a zero point length in string theory. In fact, it has been observed that such a zero point length occurs 
in string scattering processes \cite{h1, h2}. 

It has also been observed that such a zero point length occurs in other approaches to quantum gravity. This is because such a zero point length acts like an extended structure in the background geometry of spacetime. It is known that such    extended structures also occur  in loop quantum gravity \cite{z1}. In fact, such a zero point length even occurs in Asymptotically Safe Gravity \cite{asgr}  and conformally quantized quantum gravity \cite{cqqg}. So, the existence of such a minimal length seems to be an universal feature of any theory of quantum gravity \cite{a}. This can also be argued using black hole physics as any theory of quantum gravity has to be consistent with the semi-classical black hole physics. Now it is known that the energy needed to probe a region of space below Planck length is more than the energy needed to form a black hole in that region of space \cite{y1, y2}. Hence if we try to make a trans-Planckian measurement, we will form a black hole which 
will in turn prevent such a measurement. 

It has been demonstrated that such a zero point length in the background geometry of spacetime can modify the low energy 
quantum mechanical systems, as it will modify the Heisenberg uncertainty principle to a generalized uncertainty principle 
\cite{p1, p2, p4, p5}. It has also been proposed that this zero point length  can be much larger than the Planck length, 
and this can produce effects which can be measured using present experimental data \cite{s1, s2}. 
In fact, it has also been proposed that optomechanical setup can be used to test such a low energy modification 
of quantum mechanical systems from the zero point length in spacetime \cite{n1, n2, n4, n5}. It has been demonstrated that
this proposed experiment is within the reach of the current technologies, and so can be used to test the such a modification of quantum mechanics from quantum gravity. 

It may be noted that the zero point length for the geometry of spacetime can also increase in models with large extra dimensions \cite{d1, d2, d4, d5}. In such models the gravitational sector consists of closed strings, and these closed strings propagate in the higher dimensional bulk. However, the matter consists of open strings on D3-branes. Such models have been generalized to Randall-Sundrum models \cite{r1, r2}. It has been demonstrated that the low energy effective field theory of strings on a compact dimension also contains such a zero point length. This is because the T-duality for center-of-mass of the strings has been used to construct an effective path integral for such a 
system \cite{cm12, cm14, a4, a5}. This effective path integral has in turn been used to obtain modified Green’s function, with a zero point length. Thus, we can assume that the minimal length much larger than the Planck length \cite{s1, s2}, which can be measured using present experimental data, could be produced from T-duality of strings in 
such extra dimensions \cite{cm12, cm14, a4, a5}. So, generalized uncertainty can arise from these extra dimensions, and deform the Heisenberg uncertainty principle to produce a generalized uncertainty principle in four dimensions. 
 
Even though corrections to different quantum systems from generalized uncertainty have been thoroughly studied \cite{x1, x2, x4, x5}, it is important to understand if the generalized uncertainty principle is actually a quantum mechanical effect. Thus, it is important to understand how the generalized uncertainty principle will modify purely quantum effects like quantum entanglement and complementarity. These purely quantum effects are most clearly studied in quantum erasers, which can be constructed using a modified double sit experiment \cite{e1, e2, e4, e5}. 
It is known that the interference patterns disappear when we measure which of the two slits a photon has passed through in a double sit experiment. However, it is possible to eraser the information about the slit the photon has passed through using a quantum eraser. 
The interference patterns reappear after this information has been erased. These quantum erasers use quantum entanglement, and hence work on a purely quantum mechanical property. So, we will analyze how these quantum erasers get modified by the generalized uncertainty principle, and hence demonstrate that the generalized uncertainty principle actually modifies the purely quantum mechanical properties of a system. However, as the generalized uncertainty principle is based on a zero point length, which is fixed by the radius of compactification, we can argue that there is a deeper relation between geometry and quantum mechanics. 
In fact  it is  known in  AdS/CFT correspondence  the supergravity solutions in AdS are dual to conformal field theory on the boundary of that AdS spacetime \cite{ads/cft, ads/cft1}.  This is a  duality between a classical theory and a quantum theory, and for this duality to actually exists, it is important  that there exists a deeper connection  between geometry and quantum effect. So, it is important to analyze the effects of geometry of purely quantum effects. Thus, we will analyze the effect of zero point length on quantum erasers, because  zero point length is a geometric effect, and quantum erasers work on purely quantum effects.

\section{T-Duality and Zero Point Length}
It is known that Planck scale acts as the zero point length in spacetime    \cite{a, a1, a2, b1, b2}, and this   zero point length become much larger  in models with large extra dimensions \cite{d1, d2, d4, d5} or Randall-Sundrum models \cite{r1, r2}. In fact, it has been demonstrated that such a zero point would occur in  the low energy effective field theory of strings   compactified on an extra dimension \cite{cm12, cm14, a4, a5}. So, in
  this section, we will review the occurrence of zero point length in string theory, due to compactification \cite{ds1, ds2, ds3}.
Thus, for the simple case of a string in  a spacetime  with   one extra dimension $X^5$ compactified on a circle of radius $R$, we can write 
\begin{equation}
X^5(\sigma + \pi, \tau) = X^5(\sigma, \tau) + 2 \pi R w,
\end{equation}
where $w$ is the winding number. Now the mode expansion of $X^5$ can be written as 
\begin{equation}
X^5 = x^5 + 2\alpha' p^5 \tau + 2 R w \sigma + \frac{i}{2} l_s \sum_{n \neq 0} \frac{1}{n} e^{-2in\sigma} (\alpha _n^5 e^{2in\sigma} + \widetilde{\alpha}_n^5 e^{-2in\sigma}).
\end{equation}
It may be noted that as $x^5$ is periodic, that the momentum $p^5$ must be quantized 
$
p^5 =  {k}/{R}, $ with $  k\; \in\; Z, $
where $k$ is the Kaluza-Klein excitation level. The left and right movers can now be written as
\begin{eqnarray}
X_R^5(\tau - \sigma) = \frac{1}{2} (x^5 - \widetilde{x}^5) + \sqrt{2\alpha'} \alpha_0^5 (\tau - \sigma) + \frac{i}{2} l_s \sum_{n \neq 0} \frac{1}{n} \alpha_n^5 e^{-2in(\tau - \sigma)}, 
\nonumber \\ 
X_L^5(\tau + \sigma) = \frac{1}{2} (x^5 - \widetilde{x}^5) + \sqrt{2\alpha'} \widetilde{\alpha}_0^5 (\tau + \sigma) + \frac{i}{2} l_s \sum_{n \neq 0} \frac{1}{n} \alpha_n^5 e^{-2in(\tau + \sigma)}.
\end{eqnarray}
Furthermore, the zero modes for this system are given by
\begin{equation}
\sqrt{2\alpha'} \alpha_0^5 = \alpha' \frac{k}{R} - w R, \; \sqrt{2\alpha'} \widetilde{\alpha}_0^5 = \alpha' \frac{k}{R} + w R.
\end{equation}
Thus, the mass of this system can be written as 
\begin{equation}
\alpha' M^2 = \frac{\alpha' k^2}{R^2} + \frac{w^2 R^2}{\alpha'} + 2 (N + \widetilde{N} - 2).
\end{equation}
Here, we also have $N - \widetilde{N} = kw$. 
Now these equations are invariant under T-duality, which is given by 
\begin{equation}
k \; \leftrightarrow \; w, \; R \; \leftrightarrow \; \widetilde{R} = \frac{\alpha'}{R}
\end{equation}
Thus, by going from $R$ to $1/R$, the $k$ and $w$ get interchanged, and no new information is gained. So, the description of string theory below $R$ is identical to its description above $R$, and thus $R$ would fix a zero point length in this spacetime. 
 
Now we can obtain the low energy effective field theory of strings compactified on extra dimensions. This can be done using the 
 effective path integral for center-of-mass of the strings, which can be obtained from the T-duality \cite{cm12, cm14, a4, a5}. This in turn can be used to obtain a modified Green’s function, with a zero point length. 
So, using the string center of mass five-momentum $P_M \equiv (p_\mu, p_5)$, the propagation kernel can be written as \cite{cm12, cm14, a4, a5}
\begin{align}
L[x - y; \; T] = & \sum_{k = -\infty}^{\infty}\int _{z(0)=x}^{z(T)=y} \int _{x^5(0)=0}^{x^5(T)=kl_0}[Dz][Dp][Dx^5][Dp_5] \times \\
& exp\bigg[i \int_0^T d\tau (p_\mu \dot{x}^\mu + p_5 \dot{x}^5 - \frac{i}{2\mu_0} (p_\mu p^\mu + p_5 p^5)) \bigg],
\end{align}
where $\mu_0$ is a dimensional parameter. As we are treating only the center-of-mass for the string, we can neglect all of the oscillator modes, and this system can be represented by a point particle with a zero point length. Here the zero point length can be written as
\begin{equation}
 l_0 = 2 \pi R. 
\end{equation}
So, the zero point length for such a system is fixed by the compactified circle.
Now the Green's function for such a system can be written as \cite{cm12, cm14, a4, a5}
\begin{equation}
G_{reg}(p) \approx \frac{1}{(2\pi)^2} \frac{l_0 K_1(l_0 \sqrt{p^2 + m^2})}{\sqrt{p^2 + m^2}}
\end{equation}
where $K_1(l_0 \sqrt{p^2 + m^2})$ is a modified Bessel function of the second kind and $m^2 \equiv m_0^2 + {l_0^2}/{\alpha'^2}$ is the physical mass of the particle, with $m_0$ as the mass of the particle in the limit $l_0 \; \rightarrow \; 0$. 

It may be noted that these calculations can be easily generalized from $(4 + 1)$ dimensions to $(4+22)$ dimensions, and the only thing that would change is that the zero point length will be given by a more involved expression involving the geometry of the compactified $22$ dimensional manifold. 
This is because    in   toroidal compactification, with    22  compactified dimensions of radii   $R_i$, Kaluza-Klein modes $k_i$  and   winding modes $w_i$, we can still write an effective path integral   using string center of mass. In this case, the 
the T-duality of the system would relate these parameters in the system. So, using T-duality, we can still argue that there would be a zero point length in the system, and the description of this system below that zero point length would be identical to its description about that length. Thus, the main results of the paper, would not change if we analyzed the full toroidal compactification  of the string in (4+ 22) dimensions. This is because these    results     depend on the existence of such a zero point length, and they   would still hold.

\section{Low Energy Quantum Systems and Zero Point Length}
It is known that such a zero point length in the background geometry of spacetime modifies 
the Heisenberg uncertainty principle, producing a generalized uncertainty principle \cite{p1, p2, p4, p5}.  
This generalized uncertainty principle, which occurs due to a zero point length,  can be written as \cite{p1, p2, p4, p5}, 
\begin{eqnarray}
 \delta x \delta p \geq \frac{1}{2} [1 + \beta_s (\delta p)^2 + \beta_s (\langle p \rangle)^2 ], 
\end{eqnarray}
where $\beta_s$ is the parameter resulting from a zero point length. In fact, it can be demonstrated that, in the context of string theory, this parameter is related to the string length $l_s$ as \cite{a1, a2}
 \begin{equation}
 \beta_s \sim l_s^2 = \alpha'^2. 
 \end{equation}
This modification of the Heisenberg uncertainty principle also modifies the Heisenberg algebra, yielding \cite{a1, a2}
 \begin{equation}
 [x_i, p_j ] = i [\delta_{ij} + \beta_s \delta_{ij} p^k p_k + 2 \beta_s p_i p_j]. 
 \end{equation}
 Now,   in the models with large extra dimensions, the effective scale for the zero point length is  raised \cite{d1, d2, d4, d5}. In fact, it can be argued using T-duality for center-of-mass of the strings this zero point length would be of the same order as the radius of compactification of these large extra dimensions 
\cite{cm12, cm14, a4, a5}. The zero point length in the background geometry of spacetime modifies the low energy quantum mechanical systems, so the new generalized uncertainty 
principle, with this new zero point length can be written as 
\begin{eqnarray}
 \delta x \delta p \geq \frac{1}{2} [1 + 
 \beta_R (\delta p)^2 + \beta_R(\langle p \rangle)^2 ], 
\end{eqnarray}
where now the new $\beta_R$ is expressed 
 in terms of the new zero point length as 
\begin{equation}
 \beta _R \sim l_0^2 = (2 \pi R)^2. 
\end{equation}
This is because  the generalized uncertainty principle is based on the    zero point length in spacetime    \cite{a, a1, a2, b1, b2}. However, as this zero point length can increase in the    models with large extra dimensions \cite{d1, d2, d4, d5} or Randall-Sundrum models \cite{r1, r2}, the parameter in the generalized uncertainty principle will also incorporate this new zero point length in spacetime. It may be noted that  the existence of  such a large zero point length in generalized uncertainty principle has already been proposed    \cite{s1, s2} and its consequences on various low energy quantum systems have also been thoroughly studied \cite{x1, x2, x4, x5}. However, we point out here that this large zero point length in generalized uncertainty principle can emerge from the compactification of string theory on extra dimensions. This can be done by using the T-duality to argue for the existence of a zero point length much larger than the string length scale. 
So, the new deformation of the Heisenberg algebra can be written as 
 \begin{equation}
 [x_i, p_j ] = i [\delta_{ij} + \beta_R \delta_{ij} p^k p_k + 2 \beta_R p_i p_j]. 
 \end{equation}
 This deformation of the Heisenberg algebra would also deform the coordinate representation of the momentum operators to $p_i = -i (1 - \beta_R \partial^i \partial_i) $ \cite{p1, p2, p4, p5}. 
It has also been proposed that such  a deformation of the Heisenberg algebra from a zero point length, much greater than the string   length   can   be detected using optomechanical 
setup \cite{n1, n2, n4, n5}. So, such optomechanical setups can also be used to detect models with compactification, as this deformation is produced  from    compactification of strings on extra dimensions.  This can be done by analyze the corrections to the low energy systems, which would be produced by the  deformed the coordinate representation of the momentum operators. 
It is important to analyze such modifications to purely quantum effects, so we will analyze the effects of this deformation on a quantum eraser.

\section{Interference Patterns and Zero Point Length} 
It is known that the interference pattern characteristic of Young's experiment will disappear when a which-way measurement is made to discover the paths taken by the particles. However, in keeping with the complementarity principle, if the information relating to the which-way measurement is erased, the interference pattern reappears \cite{qe, qe1}. This is done in quantum eraser experiments using   entanglement \cite{e1, e2, e4, e5, e6}, and such quantum erasers have already been experimentally constructed  \cite{enta1, enta2}. 

It is important to analyze the effects of a generalized uncertainty principle on    quantum erasers. This is because such a modification of quantum mechanics is based on zero point length, which occurs    in string theory  from T-duality of the compactified   extra dimensions. This occurs due to a purely geometrical effect. So, it is not clear   how such a geometrical modification will effect the behavior of purely quantum systems,  such as quantum erasers. If it can be demonstrated that the behavior of a quantum eraser is changed due to this  zero point length, then it would demonstrate a deeper relation between quantum effects and geometry. This is important because according to AdS/CFT correspondence, there is a duality between classical geometry and   quantum theory. This can only be possible if a deep relation exists between geometry and quantum effects. 
Furthermore, the change in these systems could, in principle, be used to test the existence of such a zero point length  in string theory, if the radius of compactification is much larger than the string length scale. 

\begin{figure} [h!]
 \centering
 \includegraphics[width=175mm]{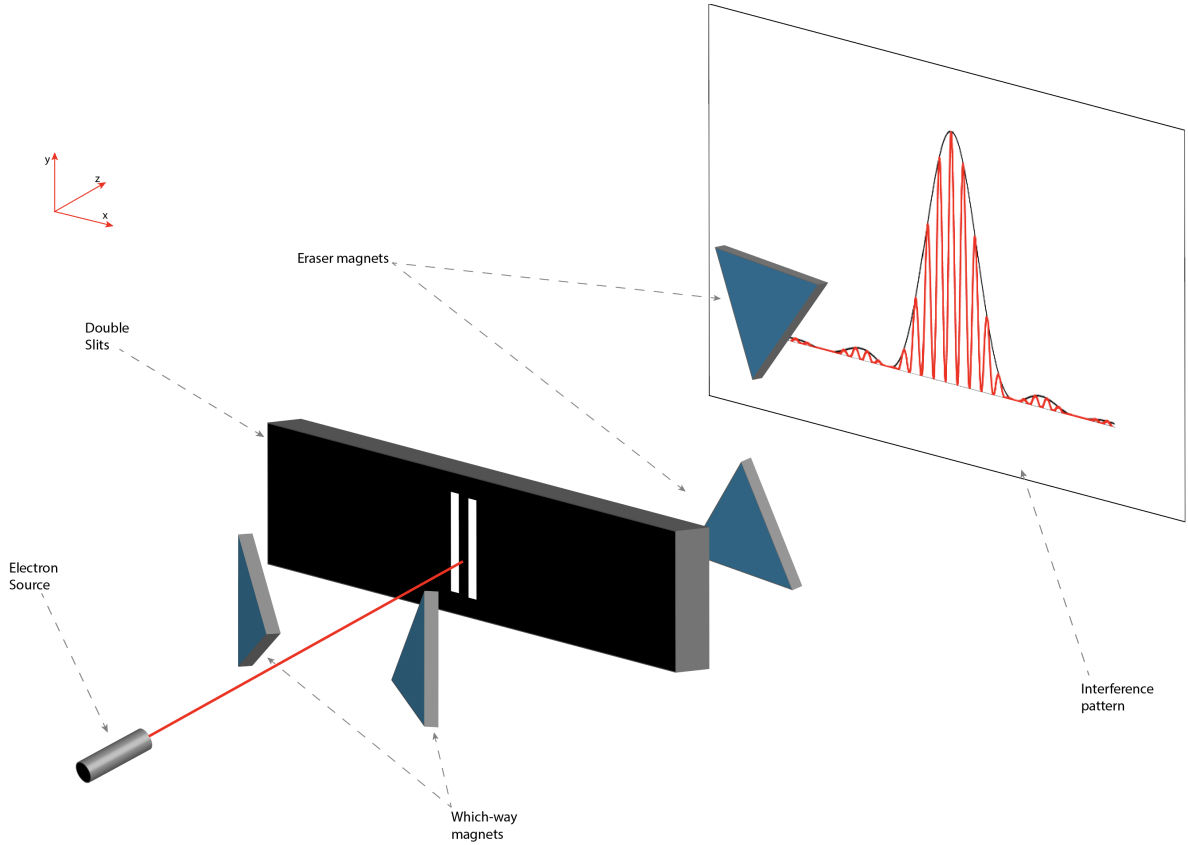}
 \caption{A modified Stern-Gerlach Quantum Eraser apparatus. Here, the particles are sent towards a double slit with spacing greater than the width of the beam. The which-way magnet separates the particles along the $x$ direction into two beams based on the $\hat{S}_x$ component of their spin allowing the beams to pass through the double slit. After the slits, the eraser magnet then again separates the two beams, this time in the $y$ direction based on the $\hat{S}_y$ component of their spin.}
 \label{ds1}
\end{figure}

Firstly we take the zero point corrected quantum mechanical state wave function of a single particle of mass $m$ and calculate the resulting fringe pattern. We assume that in the double slit apparatus the particle travels in $z$ direction with momentum $p_z$. The slits are aligned parallel to the $y$ axis, and we will only consider the interference pattern along the $x$ axis. Although we can analyze dynamics in the $y$ and $z$ axis, this setup is sufficient for analyzing the corrections due to zero point length. 
Using   quantum mechanics modified by such a zero point length, the state wave function for a particle can be written as 
\begin{equation}
\Psi_m(x)=e^{\iota k' x} +e^{-\iota k'x}
\end{equation} 
where $k'=(k+ {\beta_R^2}/{6}k^3)  = (k- g k^3)$ is the wave vector corrected by the zero point length (here we have used  $g = - \beta_R^2/6$). Now    assuming  the state wave functions to be Gaussian,  as they emerge from the slits, which are centered at $ x=+d$ and $x=-d$, we can write   
\begin{eqnarray}
 \psi_{+d}(x)=\bigg({\frac{2a}{\pi}}\bigg)^\frac{1}{4}e^{{-a(x-d)}^2},  &&
 \psi_{-d}(x)=\bigg({\frac{2a}{\pi}}\bigg)^\frac{1}{4}e^{{-a(x+d)}^2}.
\end{eqnarray}
Now we can express $\phi_i$ as a sum of $\psi_{+d}(x)$ and $\psi_{-d}(x)$ as
\begin{equation}
 \phi_i=\frac{1}{\sqrt{2}}[\psi_{+d}(x)+\psi_{-d}(x)].
\end{equation}

In the case of the Stern-Gerlach apparatus in Fig. \ref{ds1}, the which-way magnet acts to entangle the wave functions with the orthogonal spin states $ \vert+\rangle $ and $\vert-\rangle$. This is done such that if a wave packet passes through $+d$ or $-d$, it is correlated to $\vert+\rangle$ or $\vert-\rangle$, respectively.
Thus the total wave function of the single particle is the superposition of the two wave packets,  and can be written as
 \begin{equation}
  \Psi_i(x)=\frac{1}{\sqrt{2}}[\psi_{+d}(x)\vert+\rangle+\psi_{-d}(x)\vert-\rangle].
 \end{equation}
The total wave function at the screen at time $t$ is  a function of $x$, and  is given by 
\begin{equation}
 \Psi_f(x,t)=\frac{1}{\sqrt{2}}[\psi_{+df}(x,t) +\psi_{-df}(x,t).
\end{equation}

\begin{figure}[h!]
 \centering
 \includegraphics[width=80 mm]{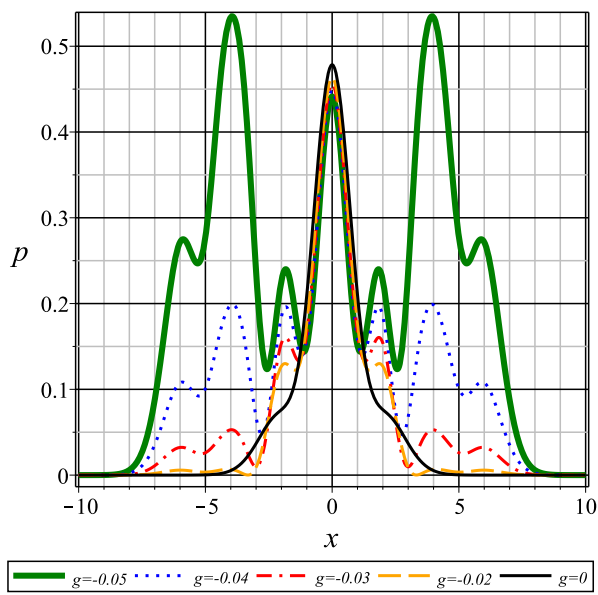}
 \caption{Typical wave function given as $P$ in terms of $x$ with $d=m=t=a=\hbar=1$ for an ordinary double slit experiment. Note that this is for a single fringe in the interference pattern.}
 \label{fig2}
\end{figure}

If we do not obtain which-way information, an interference pattern will appear at the screen
\begin{eqnarray}
 \vert\Psi_f(x,t)\vert^2&=&\frac{1}{{2}}\vert\psi_{+df}(x,t)\vert^2+\frac{1}{2} \vert\psi_{-df}(x,t)\vert^2 \nonumber \\ && +\frac{1}{{2}}\vert{\psi_{+df}}^*(x,t)\psi_{-df}(x,t)\vert +\frac{1}{{2}}\vert{\psi_{-df}}^*(x,t){\psi_{+df}}(x,t)\vert.
\end{eqnarray}
The presence of cross terms in the above suggest interference which is illustrated by Fig. \ref{fig2}.
 
Now the setup gets modified if we add the which-way magnet to the setup. So, with the added of the which-way magnet, the entangled state at the screen can be written  in terms of $\vert+\rangle$ and $\vert-\rangle$ as
\begin{equation}
 \Psi_{f_e(x,t)}=\frac{1}{\sqrt{2}}[\psi_{+df}(x,t)\vert+\rangle +\psi_{-df}(x,t)\vert-\rangle].
\end{equation}
Thus, the probability of finding the particle at position $x$ on the screen is given by 
\begin{equation}\vert \Psi_{f_e(x,t)}\vert^2=\frac{1}{{2}}\vert\psi_{+df}(x,t)\vert^2+\frac{1}{2}\vert\psi_{-df}(x,t)\vert^2\end{equation}
 
As can be seen in Fig. \ref{fig2}, if we obtain which-way information, an interference pattern will not appear on the screen. This absence is due to the vanishing of cross terms which occurs due  to orthogonality.
\begin{figure}[h!]
 \centering
\includegraphics[width=80 mm]{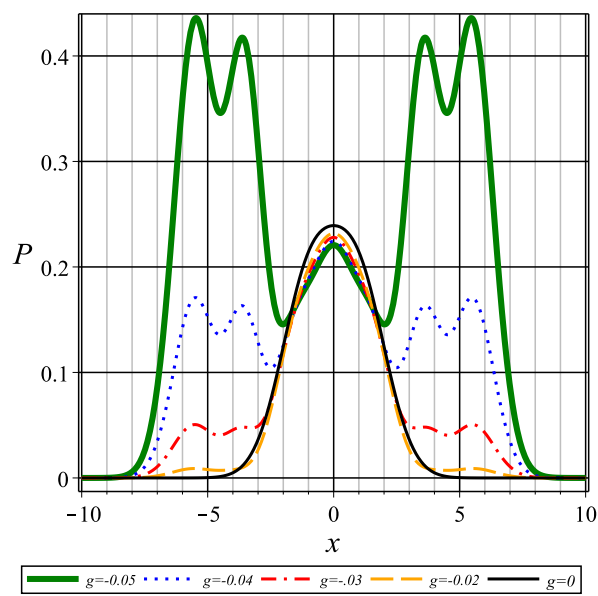}
\caption{Typical behavior of probability in terms of $x$ for $d=m=t=a=\hbar=1$ after which-way information has been measured. The result is a clear lack of interference pattern.}
 \label{fig1}
\end{figure}

\begin{figure}[h!]
 \begin{center}$
 \begin{array}{cccc}
\includegraphics[width=80 mm]{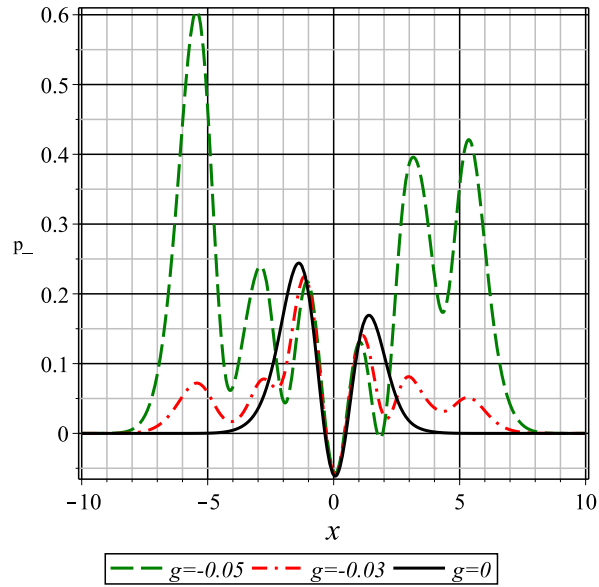}
 \end{array}$
 \end{center}
\caption{Typical behavior of $p_{-}$ in terms of $x$ for $d=m=t=a=\hbar=1$.}
 \label{fig3}
\end{figure}

Now   the eraser magnet is used to measure some non-commutating observables of $\vert+\rangle $ and $\vert-\rangle$, 
such as $\vert\uparrow\rangle $ and $\vert\downarrow\rangle$, where 
\begin{eqnarray}
 \vert+\rangle=\frac{1}{\sqrt{2}}(\vert\uparrow\rangle+\vert\downarrow\rangle), 
&&
 \vert-\rangle =\frac{1}{\sqrt{2}}(\vert\uparrow\rangle-\vert\downarrow\rangle).
\end{eqnarray}
This will erase the the which-way information, as carried by vectors $ \vert+\rangle $ and $\vert-\rangle $. Thus, we can now write 
\begin{equation}
 \Rightarrow\Psi_{f_e(x,t)}= \frac{1}{{2}}(\vert\uparrow\rangle+\vert\downarrow\rangle)\psi_{+df}(x,t)+\frac{1}{{2}}(\vert\uparrow\rangle-\vert\downarrow\rangle)\vert\psi_{-df}(x,t).
 \end{equation}
In the basis of $\vert\uparrow\rangle $ and $\vert\downarrow\rangle$ wave function can be written as 
\begin{equation}
\Psi_{f_e(x,t)}=\frac{1}{2}[\psi_{+}(x,t)+\psi_{-}(x,t)]\vert\uparrow \rangle+\frac{1}{2}[\psi_{+}(x,t)-\psi_-(x,t)]\vert\downarrow \rangle  
\end{equation}
If we measure $\vert\uparrow \rangle$ and $\vert\downarrow\rangle$, then we obtain 
 \begin{eqnarray}
  \langle\uparrow\vert \Psi_{f_e(x,t)}\rangle&=&\Psi_{f\uparrow}(x,t)=\frac{1}{{2}}[\psi_{+df}(x,t) +\psi_{-df}(x,t)]
 \nonumber \\ 
\langle\downarrow\vert \Psi_{f_e(x,t)}\rangle&=&\Psi_{f\downarrow}(x,t)=\frac{1}{{2}}[\psi_{+df}(x,t)-\psi_{-df}(x,t)] \end{eqnarray} 
It may be noted now the probabilities $p_{+}=\vert\Psi_{f\uparrow}(x,t)\vert^2$ and $p_-=\vert\Psi_{f\downarrow}(x,t)\vert^2$, can be expressed as 
 \begin{eqnarray} p_{+} &=&\frac{1}{{4}}\vert\psi_{+df}(x,t)\vert^2+\frac{1}{4} \vert\psi_{-df}(x,t)\vert^2 +\frac{1}{{2}}\vert{\psi_{+df}}^*(x,t)\psi_{-df}(x,t)\vert \nonumber \\ && +\frac{1}{{4}}\vert{\psi_{-df}}^*(x,t){\psi_{+df}}(x,t)\vert \\
 p_-&=& \frac{1}{{4}}\vert\psi_{+df}(x,t)\vert^2+\frac{1}{4} \vert\psi_{-df}(x,t)\vert^2 \nonumber \\ && -\frac{1}{{4}}\vert{\psi_{+df}}^*(x,t)\psi_{-df}(x,t)\vert -\frac{1}{{4}}\vert{\psi_{-df}}^*(x,t){\psi_{+df}}(x,t)\vert \end{eqnarray}
 Now we still have cross terms, and these cross terms suggest an interference pattern in Fig \ref{fig3}. However, the important thing to observe is that the new interference patterns, which are modified by the zero point length have reappeared. If the modification by the zero point length was not a purely quantum effect, such modified interference pattern    would not  reappear. 
 
 So, the    existence of a zero point length does change the form of such interference patterns. These interference patterns do not appear, when a way-which information is collected, even in quantum systems modified by a zero point length. However, when a way-which magnet erases this information, these new interference patterns appear again. Now what is important to note here is that the new interference patterns (which where modified by zero point length) appear again, after the way-which magnet erases the way-which information in the system. Thus, the modification 
 by the zero point length also behave as quantum modifications, which can change by not collecting the way-which information. However, as the zero point length occurs due geometry,   there seems to be a deeper connection between geometry and quantum mechanics. 
 
\section{Conclusion}
In this paper,  we have used T-duality to argue that a zero point length would occur in the low energy effective field theory of a string theory on compactified extra dimensions.
This zero point length would depend on the radius of compactification of the compact dimensions. This is because it can be demonstrated that the description of string theory below this radius is identical to its description above this length. 
Furthermore, it has also been argued that if we neglect all the oscillator modes, this zero point length would produce quantum mechanics modified by the generalized uncertainty principle. This modification of quantum mechanics would modify  various low energy quantum mechanical effects. 

Now this modified quantum mechanics based on the generalized uncertainty principle was constructed using the zero point length, which was fixed geometrically. So, it was important to analyze the behavior of the generalized uncertainty principle on purely quantum effects. As quantum erasers are constructed using purely quantum effects like entanglement. Thus the analysis of the effects of this zero point length on quantum erasers was very important.    In this paper, we have demonstrated that the behavior of these quantum erasers is modified by such a zero point length.  The   interference patterns in a double slit experiment   are changed due to this zero point length. If way-which information is collected, then these new interference patterns also disappear.  However, when the way-which information is erased by these quantum erasers, these new interference patterns reappear. Thus, the modification to the interference patterns also behave as the original interference patterns. They reappear if the way-which information is erased.

The modification to the interference patterns comes from zero point length, which in turn comes from geometry of the system. Thus, the reappearance of this change in the interference patterns in a quantum double slit experiment seems to indicate  a deeper connection between quantum effects and geometry. 
It may be noted that  in  AdS/CFT correspondence  the supergravity solutions in AdS are dual to conformal field theory on the boundary of that AdS spacetime \cite{ads/cft, ads/cft1}.  Thus, according to the AdS/CFT correspondence there exists  a relation  between    a classical theory and a quantum theory.   So, it was important to analyze the effects of geometry of purely quantum effects. Thus, we   analyzed the effect of zero point length  (a geometric effect coming from compactification) on quantum erasers (as they worked on purely quantum effects). 

It has been known for some time that quantum mechanical systems based on zero point length can have low energy consequences which can be tested experimentally \cite{s1, s2}. Furthermore, it has been proposed that this modified quantum mechanics can be tested using optomechanical systems \cite{n1, n2, n4, n5}. It would  be interesting to use a quantum double slit experiment, and quantum erasers to  test this modification of quantum mechanics. As such interference patterns can be measured with high accuracy, it could   set a bound for the radius of compactification of extra dimensions. It would also be interesting to analyze this connection between geometry and quantum erasers further. This could be done by constructing a quantum eraser in conformal field theory, then analyzing the dual to such a quantum eraser in the AdS. This could also be done with a quantum eraser modified by a zero point length using the     generalized uncertainty principle. It may be noted that the effect of the generalized uncertainty principle on AdS/CFT correspondence has already been analyzed \cite{cft1, cft2}.  So, these results can be used to analyze the   quantum erasers using the  AdS/CFT correspondence. 
 
 \section{Appendix}
As $\psi_{+d}(x) $ and $\psi_{-d}(x)$ will evolve in time and reach the screen as evolved wave packets. Thus, we can write

\begin{equation}
 \psi_{+df}(x,t)=\frac{1}{\sqrt{2}}\int_{-\infty}^{+\infty} dk\phi (k')e^{i(k'x-\frac{\ k'^2t}{2m})}.
\end{equation}

We can express $\phi(k')$ as 

\begin{equation} \phi(k')=\frac{1}{\sqrt{2}}\int_{-\infty}^{+\infty}dx ({\frac{2a}{\pi}})^\frac{1}{4} e^{{-a(x-d)}^2}e^{-ik'x}.
\end{equation}

We can also write

\begin{eqnarray}
\psi_{-df}(x,t)=\psi_{-df}(x,t)=\frac{1}{{(8a\pi^3)}^{\frac{1}{4}}}\int_{-\infty}^{+\infty}{e^{-(\frac{1}{4a}+\frac{\iota t}{2m})k^2+\iota k(1-k^2g) (x-d)-\frac{gk^4}{2a}}dk}. 
\end{eqnarray}

At this point, we make the following definitions 

\begin{eqnarray}
a_{+} \equiv \iota(x-d), &&
 \: a_{-}  \equiv  \iota(x+d),\nonumber\\
c_{+} \equiv -\iota g(x-d), &&
\: c_{-}  \equiv -\iota g(x+d),\nonumber\\
b \;  \equiv -(\frac{1}{4a}+\frac{\iota\ t}{2m}), &&
\: e  \equiv  -\frac{g}{2a}
\end{eqnarray}

Now using $b\leq0$ (as if $b>0$ then integrals diverge), and $g\ll1$, we obtain,
\begin{eqnarray}
\psi_{+df}(x,t)&=&\frac{1}{{(8a\pi^3)}^{\frac{1}{4}}}\int_{-\infty}^{+\infty}{e^{a_{+}k+bk^{2}+c_{+}k^{3}+ek^{4}}dk}\nonumber\\
&=&\frac{1}{{(8a\pi^3)}^{\frac{1}{4}}}\int_{-\infty}^{+\infty}{(1+c_{+}k^{3})(1+ek^{4})e^{a_{+}k+bk^{2}}dk},
\end{eqnarray}

So, we can write 
\begin{equation}
\psi_{+df}(x,t)=\frac{2\sqrt{\pi am}e^{-\frac{am(x-d)}{2at\iota+m}}}{{(8a\pi^3)}^{\frac{1}{4}}(2at\iota +m)^{\frac{15}{2}}}\xi
\end{equation}
where 
\begin{eqnarray}
\xi&=&64a^{6}({x}^{8}+d^{8}){g}^{2}{m}^{7}-16a^{3}({x}^{4}+d^{4})g{m}^{7}+560a^{4}{t}^{4}{m}^{3}+{m}^{7}\nonumber\\
&-&84a^{2}{}^{2}m^{5}t^{2}-128\iota a^{7}{}^{7}t^{7}+320\iota a^{6}{}^{3}t^{3}({x}^{4}+d^{4})g{m}^{4}+288a^{5}(d^{4}x^{4})g{m}^{5}{}^{2}{t}^{2}\nonumber\\
&-&112\iota a^{4}{m}^{6}t({x}^{4}+d^{4})g-768 a^{7}d^{2}gm^{3}x^{2}{}^{4}{t}^{4}+1728 a^{5}d^{2}g{m}^{5}{x}^{2}{}^{2}t^{2}\nonumber\\
&+&1920\iota a^{6}{\}^{3}{t}^{3}d^{2}g{m}^{4}{x}^{2}+448\iota a^{4}\{m}^{6}t(d^{3}x+dx^{3})g-672\iota a^{4}{m}^{6}td^{2}g{x}^{2}\nonumber\\
&+&672\iota a^{5}{}^{5}{m}^{2}{t}^{5}-280 \iota{m}^{4}a^{3}{}^{3}{t}^{3}+14\iota{m}^{6}a t+1792a^{6}(d^{6}x^{2}+d^{2}x^{6})g^{2}m^{7}\nonumber\\
&+&4480a^{6}d^{4}{g}^{2}{m}^{7}{x}^{4}-96a^{3}d^{2}g{m}^{7}x^{2}-128a^{7}(d^{4}+x^{4})gm^{3}{}^{4}t^{4}\nonumber\\
&-&1152a^{5}(d^{3}x+dx^{3})g{m}^{5}{t}^{2}+3584a^{6}d^{3}{g}^{2}{m}^{7}{x}^{5}-448a^{6}{t}^{6}m\nonumber\\
&-&1280\iota a^{6}t^{3}(d^{3}x-dx^{3})gm^{4}x+512a^{6}(d^{7}x+dx^{7}){g}^{2}{m}^{7}\nonumber\\ &+&512a^{7}(d^{3}x+dx^{3})g{m}^{3}{t}^{4}
+64a^{3}(d^{3}x+dx^{3})g{m}^{7}-3584a^{6}d^{5}g^{2}m^{7}x^{3}.
\end{eqnarray}
The 
wave function $\psi_{-d}(x)$ will evolve as
\begin{equation}\psi_{-df}(x,t)=\frac{1}{\sqrt{2}}\int_{-\infty}^{+\infty} dk \phi (k')e^{i(k'x-\frac{ k'^2t}{2m})} 
\end{equation}
Now we can write $\phi(k')$ as 
\begin{equation} 
\phi(k')=\frac{1}{\sqrt{2}}\int_{-\infty}^{+\infty}dx ({\frac{2a}{\pi}})^\frac{1}{4} e^{{-a(x+d)}^2}e^{-ik'x}
\end{equation}
We can also write 
\begin{equation}\psi_{-df}(x,t)=\psi_{-df}(x,t)=\frac{1}{{(8a\pi^3)}^{\frac{1}{4}}}\int_{-\infty}^{+\infty}{e^{-(\frac{1}{4a}+\frac{\iota t}{2m})k^2+\iota k(1-k^2g) (x+d)-\frac{gk^4}{2a}}dk}.\end{equation} 
\begin{equation}
\psi_{-df}(x,t)=\frac{2\sqrt{\pi am}e^{-\frac{am(x+d)}{2at\iota+m}}}{{(8a\pi^3)}^{\frac{1}{4}}(2at\iota +m)^{\frac{15}{2}}}\zeta
\end{equation}
where
\begin{eqnarray}
\zeta&=&64a^{6}({x}^{8}+d^{8}){g}^{2}{m}^{7}-16a^{3}({x}^{4}+d^{4})g{m}^{7}+560a^{4}{t}^{4}{m}^{3}+{m}^{7}\nonumber\\
&-&84a^{2}m^{5}t^{2}-128\iota a^{7}t^{7}+320\iota a^{6}t^{3}({x}^{4}+d^{4})g{m}^{4}+288a^{5}(d^{4}x^{4})g{m}^{5}{t}^{2}\nonumber\\
&-&112\iota a^{4}{m}^{6}t({x}^{4}+d^{4})g-768 a^{7}d^{2}gm^{3}x^{2}{t}^{4}+1728 a^{5}d^{2}g{m}^{5}{x}^{2}t^{2}\nonumber\\
&+&1920\iota a^{6}{t}^{3}d^{2}g{m}^{4}{x}^{2}+448\iota a^{4}{m}^{6}t(d^{3}x+dx^{3})g-672\iota a^{4}{m}^{6}td^{2}g{x}^{2}\nonumber\\
&+&672\iota a^{5}{m}^{2}{t}^{5}-280 \iota{m}^{4}a^{3}{t}^{3}+14\iota{m}^{6}a t+1792a^{6}(d^{6}x^{2}+d^{2}x^{6})g^{2}m^{7}\nonumber\\
&+&4480a^{6}d^{4}{g}^{2}{m}^{7}{x}^{4}-96a^{3}d^{2}g{m}^{7}x^{2}-128a^{7}(d^{4}+x^{4})gm^{3}t^{4}\nonumber\\
&+&1152a^{5}(d^{3}x+dx^{3})g{m}^{5}{t}^{2}-3584a^{6}d^{3}{g}^{2}{m}^{7}{x}^{5}+448a^{6}{t}^{6}m\nonumber\\
&+&1280\iota a^{6}t^{3}(d^{3}x+dx^{3})gm^{4}x+512a^{6}(d^{7}x+dx^{7}){g}^{2}{m}^{7}\nonumber\\ &-&512a^{7}(d^{3}x+dx^{3})g{m}^{3}{t}^{4}
-64a^{3}(d^{3}x+dx^{3})g{m}^{7}+3584a^{6}d^{5}g^{2}m^{7}x^{3}.
\end{eqnarray}

\end{document}